\documentclass[prl,aps,amsfonts,amsmath,amssymb,twocolumn,superscriptaddress]{revtex4-2}

\usepackage{graphicx}
\usepackage{bm}
\usepackage{hyperref}
\usepackage{IEEEtrantools}
\usepackage{natbib}
\usepackage{float}
\usepackage{siunitx}
\usepackage{xcolor}

\restylefloat{table}
\usepackage{dcolumn}
\usepackage{bm}
\usepackage{color}
\usepackage{braket}
\usepackage{comment}

\graphicspath{{images/}}
\usepackage[margin=1in]{geometry}
\usepackage{siunitx}

\usepackage{tabularx,ragged2e,siunitx}
\newcolumntype{Y}[1]{>{\Centeringhspace{0pt}hsize=#1hsize}X}

\begin{document}

\title{Revealing the hidden Dirac gap in a topological antiferromagnet using Floquet-Bloch manipulation}

\author{Nina Bielinski} 
\affiliation{Department of Physics, University of Illinois at Urbana–Champaign, Urbana, Illinois 61801, USA}
\affiliation{Materials Research Laboratory, University of Illinois at Urbana-Champaign, Urbana, IL 61801, USA}

\author{Rajas Chari}
\affiliation{Department of Physics, University of Illinois at Urbana–Champaign, Urbana, Illinois 61801, USA}

\author{Julian May-Mann}
\affiliation{Department of Physics, University of Illinois at Urbana–Champaign, Urbana, Illinois 61801, USA}

\author{Soyeun Kim} 
\affiliation{Department of Physics, University of Illinois at Urbana–Champaign, Urbana, Illinois 61801, USA}
\affiliation{Materials Research Laboratory, University of Illinois at Urbana-Champaign, Urbana, IL 61801, USA}

\author{Jack Zwettler} 
\affiliation{Department of Physics, University of Illinois at Urbana–Champaign, Urbana, Illinois 61801, USA}
\affiliation{Materials Research Laboratory, University of Illinois at Urbana-Champaign, Urbana, IL 61801, USA}

\author{Yujun Deng}
\affiliation{Stanford Institute for Materials and Energy Sciences, SLAC National Accelerator Laboratory, Menlo Park, CA, USA}

\author{Anuva Aishwarya}
\affiliation{Department of Physics, University of Illinois at Urbana–Champaign, Urbana, Illinois 61801, USA}
\affiliation{Materials Research Laboratory, University of Illinois at Urbana-Champaign, Urbana, IL 61801, USA}

\author{Subhajit Roychowdhury}
\affiliation{Max Planck Institute for Chemical Physics of Solids, Dresden 01187, Germany}
\affiliation{Department of Chemistry, Indian Institute of Science Education and Research Bhopal, Bhopal-462066, India}

\author{Chandra Shekhar}
\affiliation{Max Planck Institute for Chemical Physics of Solids, Dresden 01187, Germany}

\author{Makoto Hashimoto}
\affiliation{Stanford Synchrotron Radiation Lightsource, SLAC National Accelerator Laboratory, Menlo Park, CA, USA}

\author{Donghui Lu}
\affiliation{Stanford Synchrotron Radiation Lightsource, SLAC National Accelerator Laboratory, Menlo Park, CA, USA}

\author{Jiaqiang Yan}
\affiliation{Materials Science and Technology Division, Oak Ridge National Laboratory, Oak Ridge, TN, USA}

\author{Claudia Felser}
\affiliation{Max Planck Institute for Chemical Physics of Solids, Dresden 01187, Germany}

\author{Vidya Madhavan}
\affiliation{Department of Physics, University of Illinois at Urbana–Champaign, Urbana, Illinois 61801, USA}
\affiliation{Materials Research Laboratory, University of Illinois at Urbana-Champaign, Urbana, IL 61801, USA}

\author{Zhi-Xun Shen}
\affiliation{Stanford Institute for Materials and Energy Sciences, SLAC National Accelerator Laboratory, Menlo Park, CA, USA}
\affiliation{Geballe Laboratory for Advanced Materials, Stanford University, Stanford, CA}
\affiliation{Department of Physics and Applied Physics, Stanford University, Stanford, CA, USA}

\author{Taylor L. Hughes}
\affiliation{Department of Physics, University of Illinois at Urbana–Champaign, Urbana, Illinois 61801, USA}

\author{Fahad Mahmood}
\email{fahad@illinois.edu}
\affiliation{Department of Physics, University of Illinois at Urbana–Champaign, Urbana, Illinois 61801, USA}
\affiliation{Materials Research Laboratory, University of Illinois at Urbana-Champaign, Urbana, IL 61801, USA}

\maketitle

\noindent \textbf{Manipulating solids using the time-periodic drive of a laser pulse is a promising route to generate new phases of matter. Whether such `Floquet-Bloch' manipulation can be achieved in topological magnetic systems with disorder has so far been unclear. In this work, we realize Floquet-Bloch manipulation of the Dirac surface-state mass of the topological antiferromagnet (AFM) MnBi$_2$Te$_4$. Using time- and angle-resolved photoemission spectroscopy (tr-ARPES), we show that opposite helicities of mid-infrared circularly polarized light result in substantially different Dirac mass gaps in the AFM phase, despite the equilibrium Dirac cone being massless. We explain our findings in terms of a Dirac fermion with a random mass. Our results underscore Floquet-Bloch manipulation as a powerful tool for controlling topology even in the presence of disorder, and for uncovering properties of materials that may elude conventional probes.}

Floquet-Bloch manipulation refers to the time-periodic potential of an intense light pulse altering the stationary Bloch electronic states of a material. This manipulation is dictated by Floquet’s theorem, which states that a periodic Hamiltonian with driving frequency $\Omega$ has energy bands that are spaced by $\hbar\Omega$ \cite{Shirley1965}. In principle, Floquet-Bloch manipulation of materials and their topology can be specifically controlled to create new and on-demand quantum phases that cannot be achieved in equilibrium \cite{Floquet6,Kitagawa2011, Basov2017, Oka2019, rudner2020band}. The interaction between the driven and equilibrium bands can result in an effective Hamiltonian corresponding to a different phase of matter when compared to the one in equilibrium. Possibilities of such manipulation range from realizing non-equilibrium topological phases of matter, such as the Floquet-Chern insulator \cite{lindner2011floquet}, to generating Weyl semimetals \cite{Hbener2017}. So far, Floquet-Bloch engineering of materials has been experimentally achieved in 3D topological insulators \cite{gedik_science,Mahmood2016,Ito2023}, graphene \cite{GrapheneFloquet,McIver2019}, and black phosphorus \cite{Zhou2023}.

\begin{figure*}[t!]
\includegraphics[width = 1.75\columnwidth]{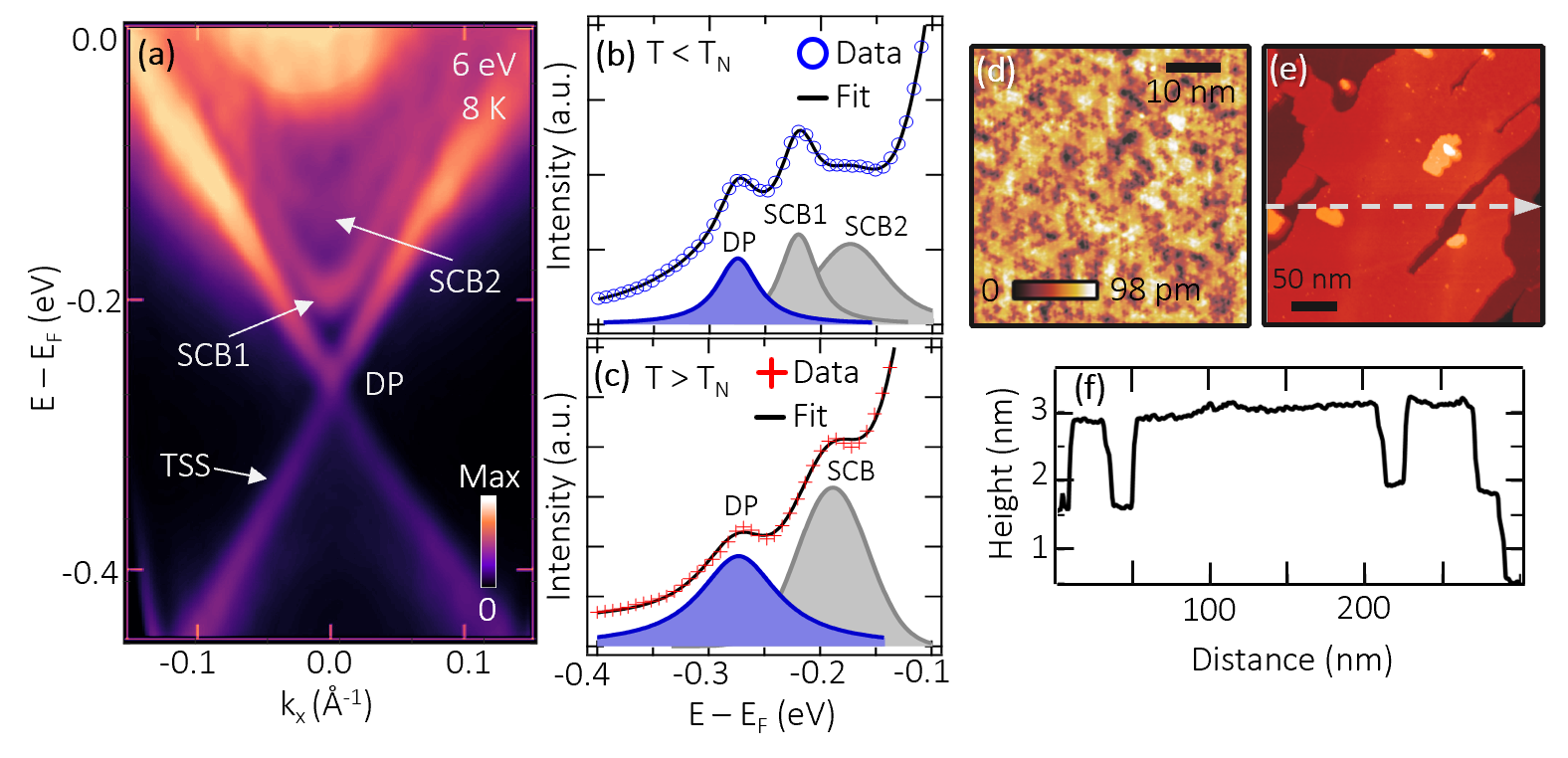}
\caption{\textbf{Equilibrium band structure and topographic maps of MnBi$_2$Te$_4$}. \textbf{(a)} The band structure of $\mathrm{MnBi_2Te_4}$ below $T_{\mathrm{N}}$ and along M-$\Gamma$-M collected with undriven 6~eV laser ARPES. The topological surface states (TSS) and the surface conduction band (SCB), which splits below $T_{\mathrm{N}}$, are labeled on the spectra. \textbf{(b, c)} Energy Distribution Curves (EDCs) along the Dirac point ($k=0$) for the laser-ARPES data in equilibrium (b) below and (c) above the Neel temperature $T_{\mathrm{N}}$. DP refers to the Dirac point. Symbols (red crosses and blue circles) indicate the raw data while solid lines are multi-peak Voigt best fits to the data. \textbf{(d)} 40~nm by 40~nm and \textbf{(e)} 300~nm by 300~nm STM topographic maps of $\mathrm{MnBi_2Te_4}$ obtained at $T$ = 300 mK indicating the presence of anti-site defects and step edges. \textbf{(f)} Topographic height vs distance along the dashed line in (e).}
\label{fig:nogap}
\end{figure*}

The past decade has also seen extensive research on 3D topological insulators that coexist with AFM order with goals of realizing, e.g., an intrinsic axion insulator phase or a high-temperature quantum anomalous Hall effect. One such intrinsic topological AFM, is $\mathrm{MnBi_2Te_4}$ \cite{NatureMBT, MBT2, MBT3, MBT4, MBT5, MBT6, MBTgap1}. Above the N\'eel temperature of $T_{\mathrm{N}}=$~25~K, $\mathrm{MnBi_2Te_4}$ is a time-reversal symmetric 3D topological insulator that hosts a single surface Dirac cone. Below $T_{\mathrm{N}}$, $\mathrm{MnBi_2Te_4}$ hosts an A-type AFM order which should induce a gap in the surface spectrum, with Ref.~\onlinecite{NatureMBT} predicting an 80~meV gap at the Dirac point of $\mathrm{MnBi_2Te_4}$. However, evidence for this gap remains controversial, with reports of either a gapless surface state \cite{MBTgap2, MBTgap3, MBTgap4, MBTgap6, MBTgap7, MBThybrid}, a large ($\sim$ 60~meV) gap \cite{NatureMBT, MBT2, MBTgap8}, or a sample dependent \cite{MBTgap5, MBTSTMgap} gap at the Dirac point. The lack of consensus on the existence of a surface Dirac mass gap is a major obstacle to fully utilizing the novel physical properties of $\mathrm{MnBi_2Te_4}$. 

In this paper, we use Floquet-Bloch manipulation to uncover the Dirac mass gap in the topological AFM $\mathrm{MnBi_2Te_4}$ below $T_{\mathrm{N}}$. Using time- and angle- resolved photoemission spectroscopy (tr-ARPES), we find that, as expected from Floquet-Bloch theory, circularly polarized light gaps out the otherwise gapless equilibrium Dirac cone in $\mathrm{MnBi_2Te_4}$ both above and below $T_{\mathrm{N}}$. For $T > T_{\mathrm{N}}$, opposite helicities of mid-IR light (right and left circular polarization) result in the same Floquet-Bloch Dirac gap. However, for $T < T_{\mathrm{N}}$, opposite helicities surprisingly produce markedly different gaps - right (left) polarized light produces a larger (smaller) gap than that expected from Floquet-Bloch manipulation of a massless Dirac Hamiltonian. The asymmetry in the Floquet-Bloch induced gaps indicates that time-reversal symmetry is indeed broken at the surface of $\mathrm{MnBi_2Te_4}$ in equilibrium. We explain the lack of an equilibrium gap by a non-uniform Dirac mass term that changes sign as a function of position on the surface. Such a system has broken time-reversal symmetry and is locally gapped away from any domain walls where the Dirac mass changes sign. However, at the domain walls, there are gapless 1D modes that make the spectrum gapless.

\begin{figure*}[t]
\includegraphics[width = 1.75\columnwidth]{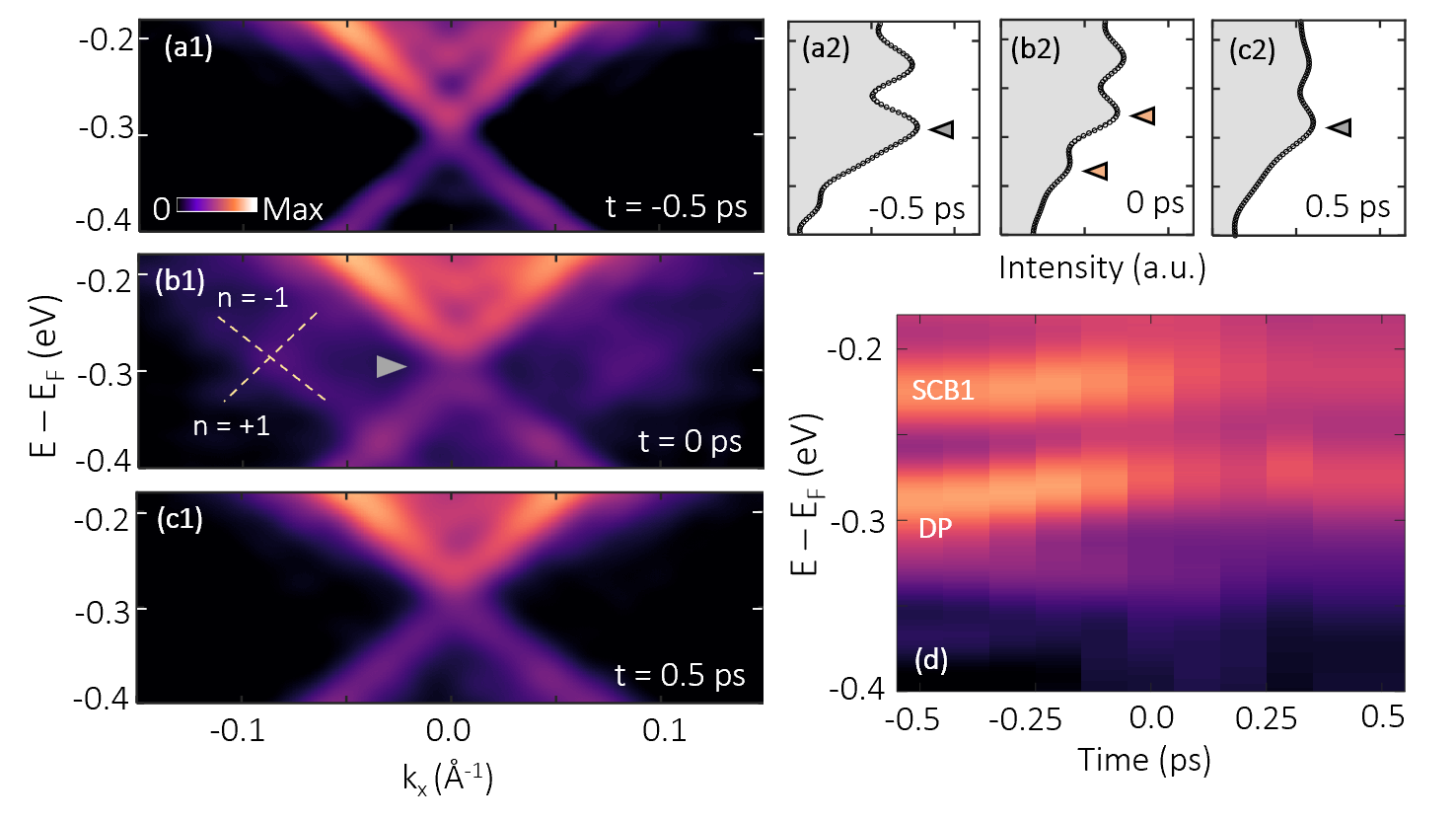}
\caption{\textbf{Floquet-Bloch gap at the Dirac point in MnBi$_2$Te$_4$ induced by circularly polarized light.} \textbf{a1-c1)} Tr-ARPES ($E - E_f$ versus $k_x$) spectra of $\mathrm{MnBi_2Te_4}$ using a right circularly polarized mid-IR ($\lambda = 8$~$\mu$m) pump for various pump-probe delay times $t$ at a temperature of $T = 8$~K. Crystal momentum $k_x$ is along the M-$\Gamma$-M direction. Sidebands of order $\pm$1 are indicated in the spectra at $t = 0$. The induced Dirac gap at $t = 0$~ps is indicated by a grey triangle in (b1). \textbf{(a2-c2)} Energy Distribution Curves (EDCs) along the Dirac point ($k = 0$) through the tr-ARPES spectra shown in (a1-c1). Triangle symbols show the positions of the peaks closest to the Dirac point. \textbf{(d)} Color plot showing the evolution of the induced Floquet-Bloch gap at the Dirac point (DP) as a function of the pump-probe delay. The color plot consists of EDCs along the Dirac point ($k = 0$) at each time point. The time-dependent data is taken in steps of 0.1~ps. SCB1 denotes the surface conduction band.}
\label{fig:snaps}
\end{figure*}

Figure~\ref{fig:nogap}a shows the equilibrium spectra of the surface of $\mathrm{MnBi_2Te_4}$ as measured using laser-ARPES (6~eV) below $T_{\mathrm{N}}$ redand without a pump pulse. The linearly dispersing topological surface states are labelled as TSS and do not show any signatures of a gap at the Dirac point. The two sharp parabolic bands above the Dira point (labelled as SCB1 and SCB2) are quasi-2D particle-like bands near the surface that are split below $T_{\mathrm{N}}$ due to the onset of AFM order as established in several ARPES studies \cite{MBTgap2,MBTgap3,MBTgap4,MBTgap9}. The absence of a Dirac gap below $T_{\mathrm{N}}$ is further indicated in the ARPES energy distribution curves (EDCs) along the Dirac point ($k_x = 0$) below and above $T_{\mathrm{N}}$ in Fig.~\ref{fig:nogap}b and Fig.~\ref{fig:nogap}c, respectively. All the equilibrium laser-ARPES measurements are done without the pump beam. As noted in Ref.~\onlinecite{MBTSTMgap}, point defects and step-edges can suppress the Dirac gap in $\mathrm{MnBi_2Te_4}$. This is relevant for our samples where the topographic STM data shown in Fig.~\ref{fig:nogap}e-g indicates the presence of substantial surface inhomogeneity from anti-site defects and step-edges. Moreover, since the surface Dirac mass arises from the time-reversal symmetry breaking AFM order, crystal defects--like step edges and antisite defects--take on a magnetic character and naturally bind in-gap states. For example, since the AFM ordering vector is normal to the surface, step edges lead to domain walls of the surface Dirac mass term, which, in turn, will bind one-dimensional chiral states.

\begin{figure*}[t!]
\includegraphics[width = 1.9\columnwidth]{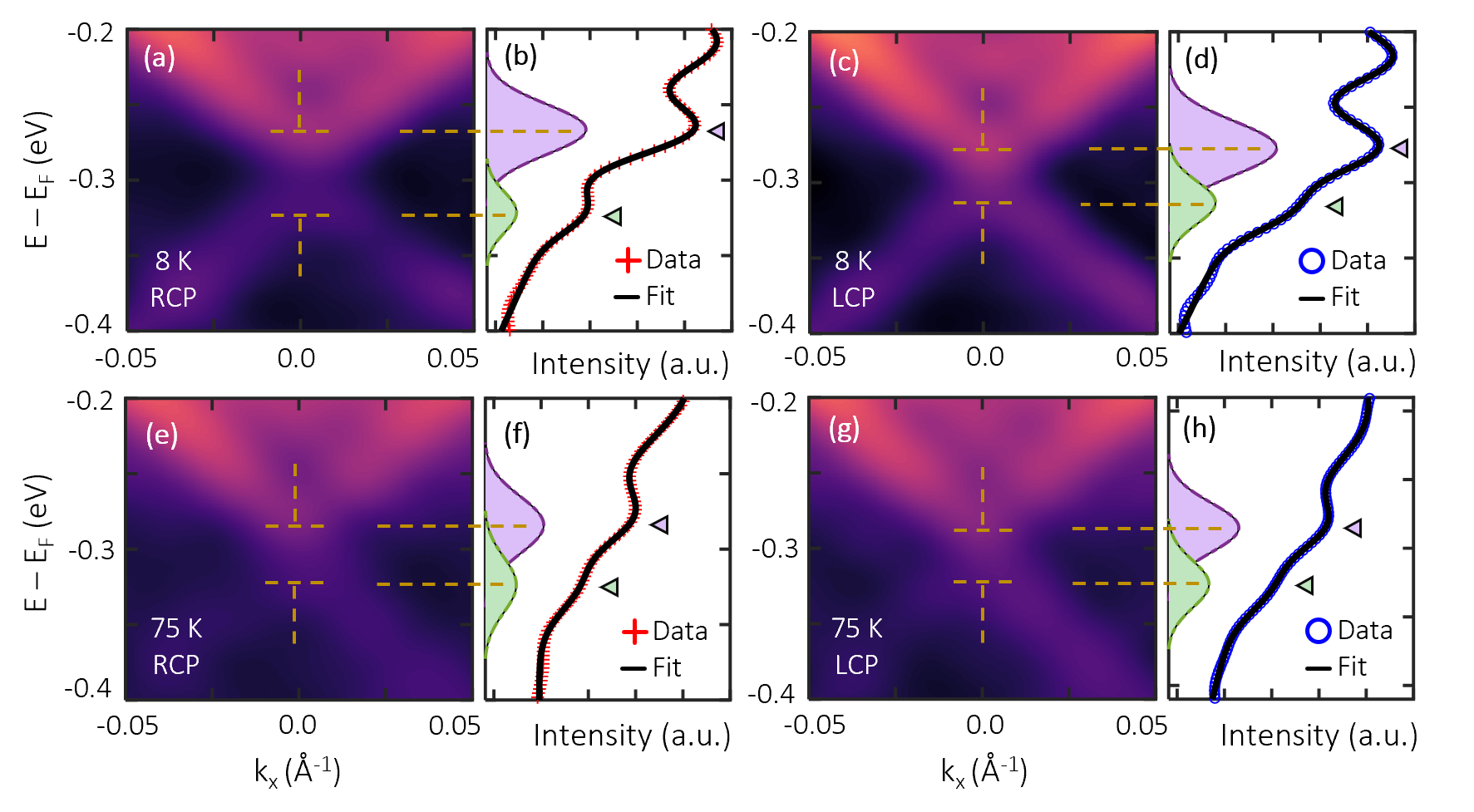}
\caption{\textbf{Opposite helicities induce different (same) Dirac gaps below (above) the Neel temperature $T_{\mathrm{N}}$.} Tr-ARPES spectra ($E-E_f$ versus $k_x$) at $t = 0$~ps for right (RCP) and left (LCP) circularly polarized pump at a temperature of \textbf{(a,c)} $T = 8 $~K ($< T_{\mathrm{N}}$) and \textbf{(e,g)} $T = 75$~K ($> T_{\mathrm{N}}$). \textbf{(b,d,f,h)} Energy Distribution Curves (EDCs) along the Dirac point ($k = 0$) through the tr-ARPES spectra shown in (a,c,e,g). Symbols (red crosses and blue circles) indicate the raw data while solid lines are multi-peak Gaussian best fit to the data. Dashed lines indicate peaks from which the Dirac gap is determined.}
\label{fig:hel-dep}
\end{figure*}

We next apply Floquet-Bloch manipulation on the surface of $\mathrm{MnBi_2Te_4}$ to alter the gapless Dirac point. Tr-ARPES measurements were taken using circularly polarized mid-IR 155~meV ($\lambda = 8 \mu m$) pulses as the pump and 6~eV pulses as the probe. We select an energy of 155~meV for the pump since it is below the 200~meV bulk band gap of  $\mathrm{MnBi_2Te_4}$ \cite{NatureMBT}, and so we limit excitations across the bulk band gap which can diminish the coherent interaction between the pump pulse and the Dirac surface states. This approach is similar to the one used in previous Floquet-Bloch experiments on 3D topological insualtors \cite{gedik_science, Mahmood2016}. Fig.~\ref{fig:snaps}a1-c1 show snapshots of the tr-ARPES spectra near the Dirac point before, during, and after the arrival of a right-circularly polarized pump at a temperature of $8$~K. Replicas of the original Dirac cone appear when the pump and probe pulses overlap in time as indicated in Fig.~\ref{fig:snaps}b1. These replicas correspond to photon-dressed states of the surface Dirac cone, as discussed in several tr-ARPES experiments on 3D TIs \cite{gedik_science,Mahmood2016,Ito2023}. We note that the replica bands observed in our work (labelled as n~=~-1 and n~=~1 in Fig.~\ref{fig:snaps}b1) are due to photon-dressing of both the initial (Floquet-Bloch) and final (Volkov) states involved in photoemission, rather than just the final states (SI). More importantly, as shown in Fig.~\ref{fig:snaps}b1, the spectra at time-zero ($t=0$) shows a clear signature of a gap opening at the Dirac point when compared to the spectra at $t = -0.5$~ps (before the arrival of the pump pulse). In addition, EDCs taken along $k_x = 0$ also show an onset of a double-peak structure at the Dirac point at $t = 0$~ps (Fig.2b2), when compared with the spectra at $t = -0.5$~ps (Fig.2a2) and at $t = 0.5$~ps (Fig.2c2). The opening of the gap by the pump pulse is also apparent in a color plot of the measured photoemission intensity around the Dirac point ($k_x$ = 0) as a function of energy and time in Fig.~\ref{fig:snaps}d. As can be seen, the Dirac point (labelled as DP) for $t = -0.5$~ps splits into two parts when the pump and probe overlap (i.e., around $t = 0$~ps), and there is a clear suppression of intensity in the middle of the two parts at $t = 0$~ps. In contrast, the band labelled as SCB1, a known conduction band near the surface for MnBi$_2$Te$_4$ does not show any splitting but only moves up in energy slightly.

The manipulation of the Dirac gap with circularly polarized light can be understood in the context of an effective Floquet Hamiltonian \cite{Floquet6,Kitagawa2011}. Note that this approach is valid assuming that electron relaxation and interaction effects are negligible, which is reasonable given that this work focuses on the transient band-structure at time-zero rather than at later times. In equilibrium, the surface of a topological antiferromagnet such as $\mathrm{MnBi_2Te_4}$ can be described with the massive Dirac equation: 
\begin{equation}\begin{split}
\mathcal{H}_{\text{Dirac}}(\vec{k}) = \hbar v_f k_x \sigma^x + \hbar v_f k_y \sigma^y + m_0 \sigma^z,
\label{eq:EquibDirac}\end{split}\end{equation} 
where $m_0$ is the equilibrium mass, and $v_f$ is the Dirac velocity. Circularly polarized light minimally couples to the surface Dirac fermions via $(k_x, k_y)\rightarrow (k_x + eA_x, k_y + eA_y)$, where $(A_x,A_y) = (\mathcal{A} \cos(\Omega t), \mathcal{A} \sin(\Omega t))$. Here $\mathcal{A} = E/\Omega$, $E$ is the electric field and $\Omega$ is the frequency of the circularly polarized light. Different signs of $\Omega$ correspond to right and left polarization.

Using the Floquet-Magnus expansion \cite{Floquet6,Kitagawa2011,bukov2015universal}, the effective Hamiltonian for the surface Dirac fermion coupled to circularly polarized light is then given by 
\begin{equation}\begin{split}
\mathcal{H}_{\text{Dirac}}(\vec{k}) = &\hbar v_f k_x \sigma^x + \hbar v_f k_y \sigma^y \\ &+ \left[m_0 + \frac{(ev_f \mathcal{A})^2}{\hbar \Omega} \right] \sigma^z.
\label{eq:DrivenDirac}\end{split}\end{equation}
The effective Hamiltonian implies a striking difference between Floquet-Bloch manipulation of a massless and massive Dirac system. For a massless Dirac cone ($m_0 = 0$ in equilibrium), circularly polarized light results in a Dirac mass gap of $\Delta = 2|\frac{(ev_f \mathcal{A})^2}{\hbar \Omega}|$, which is the same for left ($\Omega < 0$) and right ($\Omega > 0$) polarized light. For a massive Dirac cone ($m_0 \neq 0$ in equilibrium)  the gap in Eq.~\ref{eq:DrivenDirac} is $2|m_0 + \frac{(ev_f \mathcal{A})^2}{\hbar\Omega}|$ which depends on the relative signs of $m_0$ and $\Omega$. If $m_0$ and $\Omega$ share the same sign, the two mass terms cooperate, increasing the total mass of the Dirac fermion. If $m$ and $\Omega$ have opposite signs, the two mass terms compete, and the application of circularly polarized light will shrink the surface gap ($|m_0| > |\frac{(ev_f \mathcal{A})^2}{\hbar\Omega}|$) before closing ($|m_0| = |\frac{(ev_f \mathcal{A})^2}{\hbar\Omega}|$) and then reopening ($|m_0| < |\frac{(ev_f \mathcal{A})^2}{\hbar\Omega}|$). Overall, this indicates that changing the helicity of circularly polarized light (and keeping the intensity constant) should change the gap of a driven topological antiferromagnet below $T_{\mathrm{N}}$. 

\begin{figure*}[ht!]
\includegraphics[width = 1.8\columnwidth]{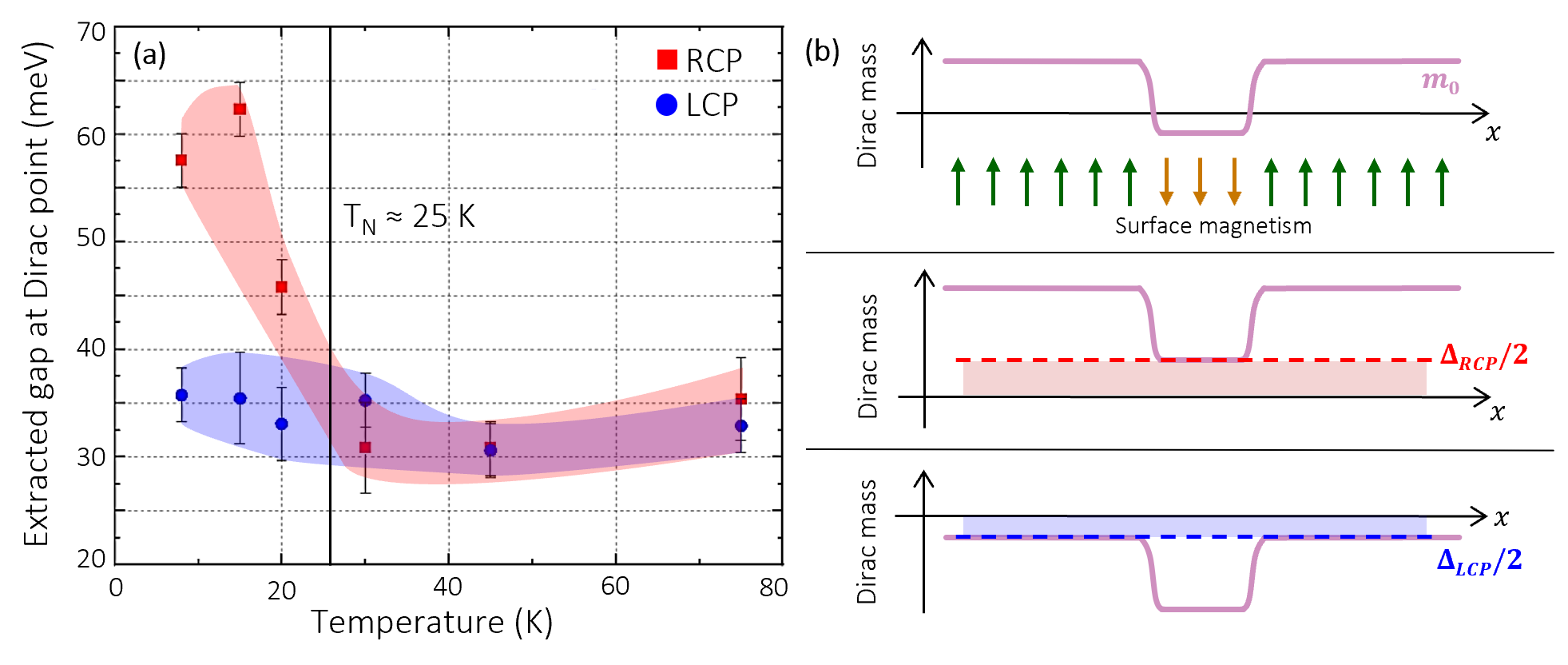}
\caption{\textbf{Temperature dependence of the Floquet-Bloch induced Dirac gap} \textbf{(a)} Floquet-Bloch induced gap as a function of temperature for both right (RCP) and left (LCP) circularly polarized light. Shaded regions are a guide to the eye. Error bars represent the 95$\%$ confidence interval (2~s.d.) in extracting the gap from the fitting parameters. \textbf{(b)} Model representing the non-uniform surface Dirac mass term of $\mathrm{MnBi_2Te_4}$ in equilibrium (top panel), when driven with RCP light (middle) and when driven with LCP light (bottom).}
\label{fig:T-dep}
\end{figure*}

However, observing this helicity dependent behavior would require a massive Dirac cone in equilibrium - which is not the case for our $\mathrm{MnBi_2Te_4}$ samples (Fig.~\ref{fig:nogap}). If the surface of $\mathrm{MnBi_2Te_4}$ is truly gapless, right and left circularly polarized pumping should produce the same Dirac gap in our measurements regardless of temperature. To check this expectation and to quantify the measured Dirac gap, we next focus on the tr-ARPES spectra taken at $t = 0$ near the Dirac point for temperatures above and below $T_{\mathrm{N}}$. Figure~\ref{fig:hel-dep}e-h show the tr-ARPES spectra at $75$~K for right and left circularly polarized light respectively. Both polarizations result in very similar Dirac gaps as is clear from fits to the energy distribution curves (EDCs) along the Dirac point (Fig.~\ref{fig:hel-dep}f and h). Details of the fitting procedure for the EDCs are discussed in the SI. Note that the measured Floquet-Bloch induced Dirac gap of $\sim 35$ meV is in agreement with the theoretical predicted value (Eq.~\ref{eq:DrivenDirac}) of $\sim 30$ meV for $m = 0$ (SI).

Surprisingly, the tr-ARPES spectra at the Dirac point for $T = 8$~K are noticeably different for right and left circularly polarized pump, as shown in Fig.~\ref{fig:hel-dep}a to d. The gap induced by right polarized light is about twice that induced by left polarized light as shown in the corresponding EDCs (Fig.~\ref{fig:hel-dep}b and d). As noted above, this difference is expected to occur if the equilibrium Dirac mass is non-zero. Given the parameters of our drive (Methods and SI), the difference in the induced mass would suggest an equilibrium Dirac mass of about $ m = 20 $ meV), above the energy resolution of both our laser-ARPES and synchrotron ARPES measurements. However, a Dirac gap is not observed in any of our equilibrium measurements across various sample cleaves and positions (Fig.~S2 in SI). 

To study this further, we performed tr-ARPES measurements across a range of temperatures both above and below $T_{\mathrm{N}}$. Figure~\ref{fig:T-dep}a shows the induced Dirac mass gap as a function of temperature for right and left polarized light. As can be seen, the induced gap is different for the two polarizations below $T_{\mathrm{N}},$ but it is the same above $T_{\mathrm{N}}$. We note here that there is negligible change in the optical absorption above and below $T_{\mathrm{N}}$ at the pump wavelength of 8~$\mu$m in $\mathrm{MnBi_2Te_4}$ \cite{Xu2021}, and so the difference observed in the induced gap cannot be attributed to changes in the optical absorption. This is clear evidence that time-reversal symmetry is broken at the surface of $\mathrm{MnBi_2Te_4}$ below $T_{\mathrm{N}}$, not just in the bulk. 

The absence of an observed Dirac gap in equilibrium measurements and the time-reversal symmetry breaking observed in the Floquet-Bloch experiments, can be reconciled by the existence of a non-uniform mass at the surface of $\mathrm{MnBi_2Te_4}$ (i.e., taking $m_0$ to be a function of position in Eq.~\ref{eq:EquibDirac}). The non-uniformity of the mass term occurs because of site defects and magnetic domain walls associated with step edges. In particular, site defects can lead to local regions where the Dirac mass vanishes \cite{liu2022visualizing}, while magnetic domain walls induce Dirac mass domain walls, which are known to bind chiral modes \cite{shen2012topological}. Application of circularly polarized light induces a second constant mass term (Eq.~\ref{eq:DrivenDirac}), that shifts the Dirac mass profile up or down. Provided the amplitude of the photo-induced mass is large enough, shifting the Dirac mass profile will remove all domain walls and zero-gap regions, and the accompanying mid-gap modes, as illustrated in Fig.~\ref{fig:T-dep}b. We note here that disorder due to site defects and magnetic domain walls is likely not so strong as to produce strong localization, and so the observed spectral features in ARPES experiments (e.g.~Fig.1) remain sharp.

To illustrate how the non-zero average of the Dirac mass leads to the observed asymmetry between right and left circularly polarized light, we model the surface as a collection of regions, where the surface mass is roughly constant. From the combined effects of magnetic domain walls and site defects, the mass in these regions can be either positive, negative, or zero. Provided the regions are significantly large, the measured mass gap can be approximated as $\min(2|m_0(\bm{r}) + m_c|)$, where $m_0$ is the non-uniform equilibrium mass, and $m_c$ is the spatially uniform light-induced mass term. As in the Floquet-Magnus expansion, Eq.~\ref{eq:DrivenDirac}, $m_c>0$ ($m_c<0$) for right (left) circularly polarized light. If we take the magnitude of the light-induced mass to be larger than the maximum magnitude of the equilibrium mass, $|m_c| > \max(|m_0|)$ (suppressing the dependence of $m_0$ on position $\bm{r}$), then a straightforward calculation gives $|\Delta_{\text{RCP}} - \Delta_{\text{LCP}}| = 2|\bar{m}_{0}|$, where $\Delta_{\text{RCP}/\text{LCP}}$ is the measured mass gap for right and left circularly polarized light, and $\bar{m}_{0} \equiv \max(m_0) + \min(m_0)$. We therefore find that if the surface mass is larger in magnitude for the positive mass domains than the negative mass domains, right and left circularly polarized light will produce unequal gaps. From this, we conclude that a generic non-uniform mass term will lead to the observed time-reversal symmetry breaking, since there is no symmetry or other constraint that would cause the magnitude of the mass to be equal in the positive and negative domains. The surface of $\mathrm{MnBi_2Te_4}$ is likely more complicated than the collection of uniform mass regions we considered here. Nevertheless, this simple picture is sufficient to establish how a non-uniform surface mass term can lead to the observed difference in Dirac gaps for right and left circularly polarized light. 

To conclude, we point out that a Dirac fermion with a random mass can be mapped onto the Chalker-Coddington network model \cite{Chalker1988,Ho1996,Song2021}. This model was originally used to describe transitions between integer quantum Hall states, where a disordered potential creates a network of chiral modes that move along equipotential lines. On the surface of $\mathrm{MnBi_2Te_4}$, a similar network of chiral modes move along magnetic domain walls, resulting in the same physical picture. Circularly polarized light can be used to control the mass of the $\mathrm{MnBi_2Te_4}$ surface states, which, in turn, determines how much of the chiral modes on different domain walls overlap with one another. Hence, our results suggest that Floquet-Bloch manipulation using circularly polarized light can serve as a tuning knob for the tunneling parameters in the Chalker-Coddington description. It should therefore be possible to observe the universal behavior of the Chalker-Coddington network model in driven $\mathrm{MnBi_2Te_4}$ and similar topological AFM materials. 

\vspace{4mm}
\noindent\textbf{Acknowledgements:} We thank Peter Abbamonte, Peter Armitage, Barry Bradlyn, and Ken Burch for insightful discussions.\\
This work was supported by the Quantum Sensing and Quantum Materials, an Energy Frontier Research Center funded by the U.S. Department of Energy (DOE), Office of Science, Basic Energy Sciences (BES), under Award No.DE-SC0021238. F.M. acknowledges support from the EPiQS program of the Gordon and Betty Moore Foundation, Grant GBMF11069. N.B. acknowledges support from the Illinois Materials Research Science and Engineering Center, supported by the National Science Foundation MRSEC program under NSF Award No.\ DMR-1720633. Y.D. acknowledges support from the Bloch Fellowship in quantum science and engineering from Stanford Q-FARM. Work at SIMES and ORNL was supported by the U.S. Department of Energy, Office of Science, Basic Energy Sciences, Materials Sciences and Engineering Division.\\

\vspace{4mm}
\noindent\textbf{Author contributions:} N.B., S.K., J.Z. and F.M. performed the tr-ARPES experiments and the corresponding data analysis. R.C., J.M-M., and T.L.H. developed the theoretical understanding. Y.D., M.H., D.L., and Z-X.S provided crucial insights into understanding of the data based on synchrotron ARPES experiments. A.A. and V.M. performed the STM experiments. S.R., C.S. and C.F. synthesized the MnBi$_2$Te$_4$ samples for the tr-ARPES and STM experiments. J.Y. synthesized the samples for the synchrotron ARPES experiments. N.B., J.M-M., and F.M. wrote the manuscript with input from all the authors. F.M. conceived and supervised this project.

\vspace{4mm}
\noindent\textbf{Competing interests:} The authors declare no competing interests.

\vspace{4mm}
\noindent\textbf{Methods}

\vspace{2mm}

\noindent\textit{tr-ARPES:} tr-ARPES measurements were performed using a 6 eV probe beam and mid-IR pump beam ($\lambda = 8\mu$m) generated by a Yb:KGW amplified fiber laser. $\mathrm{MnBi_2Te_4}$ samples were cleaved \textit{in situ} at a base pressure of under 7 x $10^{-11}$ mbar. ARPES measurements were performed using a Scienta Omicron hemispherical analyzer with deflection angle mapping capability (DA-30L). This system has energy and momentum resolutions better than 10~meV and 0.002~Å$^{-1}$, respectively.

\vspace{2mm}

\noindent\textit{STM measurements:} The scanning tunneling microscopy (STM) measurements were performed on $\mathrm{MnBi_2Te_4}$ single crystals cleaved at 90~K in UHV and immediately inserted into the instrument. The data was obtained on a commercial Unisoku STM operating at 300 mK. dI/dV spectra were collected using a standard lock-in technique at a frequency of 913 Hz.

\vspace{2mm}

\noindent\textit{Sample synthesis and characterization:} MnBi$_2$Te$_4$ single crystals were grown using the Bi$_2$Te$_3$ flux method \cite{Yan2019}. As-purchased elemental manganese (99.999\%, Alfa Aesar), bismuth (99.997\%, Alfa Aesar), and tellurium (99.9999\%, Alfa Aesar) were mixed in the molar ratio of Mn:Bi:Te of 1:10:16 in an argon-filled glove box. All of the elements were loaded into an alumina crucible which was vacuum sealed in a quartz tube under (10$^{-5}$ Torr). The tube was heated to 950$^\circ\mathrm{C}$ for 9 hours, then held for 24 hours before being progressively cooled to 588$^\circ\mathrm{C}$ for 100 hours. After centrifugation at 588$^\circ\mathrm{C}$ to remove excess Bi$_2$Te$_3$, the crystals were recovered. The single crystal has a typical dimensions of 3×3×0.4 mm$^3$. 

\vspace{2mm}

\noindent\textit{Determination of the Dirac gaps:} The Floquet-Bloch induced Dirac point gaps shown in Fig.~3 and Fig.~4 are determined by EDCs at ($k_x$,$k_y$) = (0,0) through the tr-ARPES spectra obtained at $t = 0$. Each EDC is fitted with three multi-peak Voigt functions with a linear background. One peak corresponds to the surface conduction band (SCB1) as shown in Fig.~1a while the Dirac gap is determined by the distance between the other two peaks (shaded peaks in Fig.~3). The specific choice of the peak functions does not affect the obtained gap size in any significant way since we are only interested in the distance between the peaks.

\vspace{4mm}
\noindent\textbf{Data availability:} Data are available from the corresponding author upon reasonable request. 

\newpage
\bibliographystyle{apsrev}
\bibliography{Bibiliography}
\newpage

\end{document}